\documentclass[11pt,twoside,a4paper]{article}
\usepackage{epsfig}
\usepackage{graphicx}
\usepackage{srt_style}
\begin{document}
\baselineskip .5cm
%
\title{Current and Future Developments in Deep, Wide--field VLBI
  Continuum Surveys}

\author{M.A. Garrett}

\affil{JIVE Institute for VLBI in Europe, Postbus 2, 7990~AA
  Dwingeloo, The Netherlands.}

%
\begin{abstract}
  I review the current status of deep, wide-field VLBI continuum
  surveys. I also discuss anticipated short and long-term improvements
  in sensitivity (e.g. the eEVN), and the science these developments
  will enable.
\end{abstract}

\section{Introduction}

In the last few years, great strides have been made towards
understanding the nature of the faint sub-mJy and microJy radio source
population. Most of these studies have employed connected arrays, such
as the VLA {\it e.g.} Richards et al. 1998, Richards 2000; MERLIN {\it
  e.g.} Muxlow et al. 1999; ATCA {\it e.g.} Norris et al. 2001 and the
WSRT {\it e.g.} Garrett et al. 2001). These observations typically
integrate for many days on relatively small patches of sky, targeting
well known ``deep field'' regions, rich in complimentary observations
at sub-mm, mid-IR, near-IR, optical and x-ray wavelengths.  Such radio
observations reach r.m.s.  noise levels $\sim$~few microJy/beam,
encountering source densities of $\sim 1$ per square arcmin at 1.4 GHz.

Until recently the idea that VLBI observations could detect even one
solitary, compact radio source in these narrow, radio-quiet, deep field
regions would have been tantamount to a declaration of professional
suicide. However, such observations are now possible due to: (i)
increasing individual telescope sensitivity coupled with sustained,
high data rate recording systems, (ii) the simultaneous employment of
both phase-referencing and wide-field imaging techniques, and (iii) 
the availability of more capable correlators, matched by equally impressive
desktop data analysis facilities. 

In this paper I will review the current status of deep, wide-field,
VLBI continuum imaging - focusing on the technique and the science it
enables. I will also look forward to future developments (e.g.
$e$EVN) -- these promise to completely transform our current capabilities,
in a way that will open-up entirely new areas of VLBI research. But
first I begin with a summary of the properties of the faint, sub-mJy
and microJy radio sky.
\clearpage 

\begin{figure}[h]
\vspace{12.5cm}  
\includegraphics{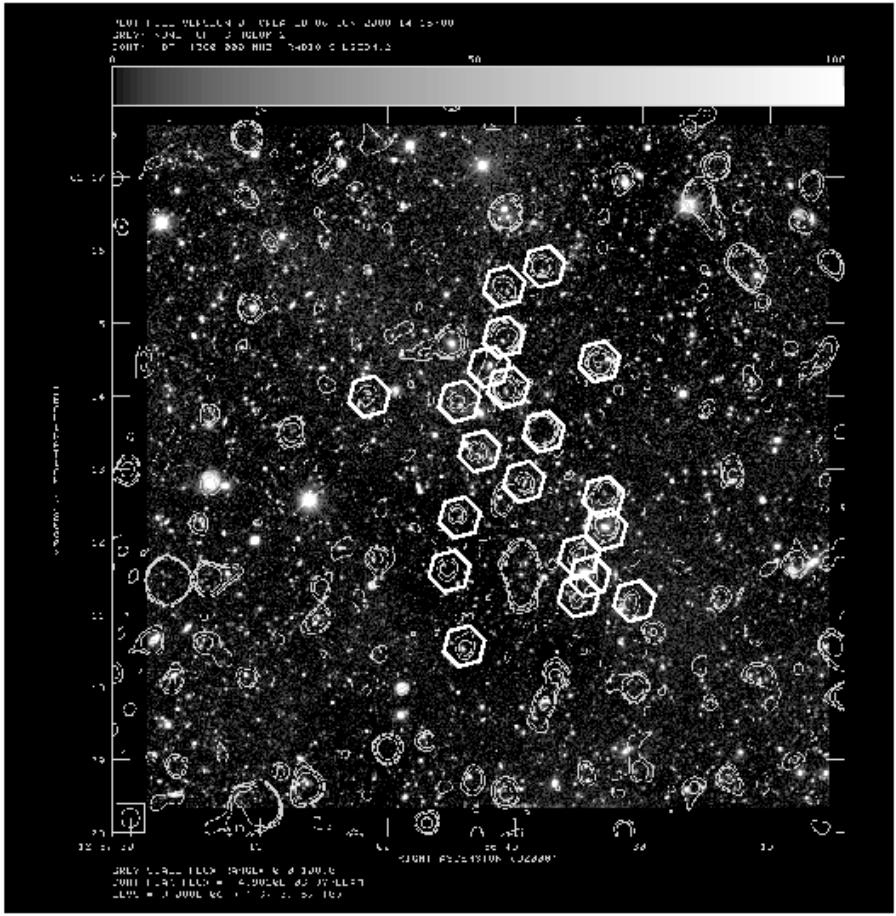}
\caption{A deep WSRT 1.4~GHz radio contour map superimposed 
  upon a CFHT optical image (Barger et al. 1998) of the Hubble Deep
  Field and Flanking Field region. Mid-IR $15\mu$m ISO sources ($>
  4\sigma$ detections that coincide with WSRT radio sources) are
  represented by open hexagonal symbols. The ISO observations (Aussel
  et al. 1999) only cover the inner part of the full WSRT field of
  view, nevertheless the correlation between ISO and WSRT radio sources
  is striking. }
\label{iso_wsrt} 
\end{figure}

\section{Faint Radio Sources} 

\begin{figure}[h]
\vspace{8.5cm}  
\includegraphics{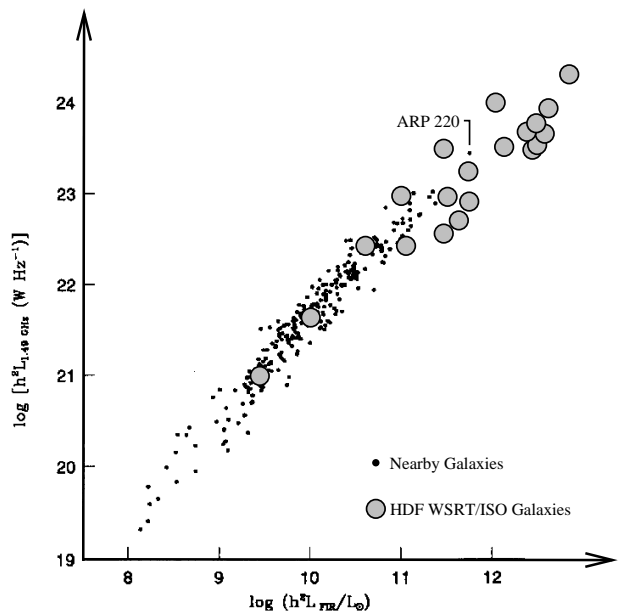}
\caption{A logarithmic plot of the FIR vs Radio Luminosity 
  for local galaxies (Condon 1992) with the addition of the ISO/WSRT
  extension to fainter but higher luminosity/higher redshift sources
  shown by the larger filled circles (Garrett 2002). The additional
  ISO/WSRT sample covers a range of redshift, up to $z \sim 1.3$. The
  most luminous sources are also the most distant, with implied SFRs
  an order of magnitude greater than Arp 220.}
\label{fir-rc} 
\end{figure}

Fig.~1 presents a deep 1.4 GHz WSRT image (Garrett et al. 2000a) of the
Hubble Deep Field (HDF -- Williams et al. 1996).  One of the simplest
but most striking observations is the close correspondence between
faint WSRT 1.4 GHz detections and ISO $15\mu$m detections. In
particular, the vast majority of these distant sources that are common
to both the WSRT and ISO appear to follow the well-known FIR-radio
correlation (Garrett 2002).  This FIR-radio correlation was first
observed with respect to ``normal'' star-forming galaxies in the local
Universe (e.g. Condon 1992). Its extension to these much more luminous
and more distant systems in the HDF (the WSRT/ISO sample has a median
redshift of 0.7) is convincing evidence that the majority of the faint
radio source population are identified as star forming galaxies with
Star Formation Rates (SFR) often well in excess of 100~M$_{\odot}$/yr. This
observation (see Fig.~\ref{fir-rc}) is backed up (Richards et al. 1998,
Muxlow et al. 1999) by the measured angular size of the radio sources,
their radio morphology/steep radio spectra, and their optical
morphology (the latter often showing evidence for galaxy interactions).

A picture is thus developing of a radio sky that literally ``lights
up'' at microJy flux density levels, with the bulk of the radio
emission being directly related to massive star-formation i.e.
supernovae events and the global acceleration of cosmic ray electrons
via propagating shocks that are associated with these events. Note,
however, that these conclusions are dominated by the faintest (and
most numerous) microJy radio sources, in particular the AGN
fraction is observed to increase rapidly at higher (sub-mJy) flux
density limits. In addition, there is a significant fraction of these
faint radio sources for which little is known, these are the optically
faint radio source population -- believed to be high redshift galaxies
in which the dust content is large enough to completely obscure the IR,
optical, uv and even soft x-ray emission. There seems to be a correspondence
between this population and the faint SCUBA source population (e.g.
Hughes et al. 1998). The latter are now considered to be responsible
for the vast bulk of the star formation in the early Universe but their
detailed properties remain largely unknown. In particular, it is
unclear whether the radio and FIR emission arise via AGN, star formation
activity or some combination of both phenomena.

\begin{figure}[h]
\vspace{13cm}  
\includegraphics{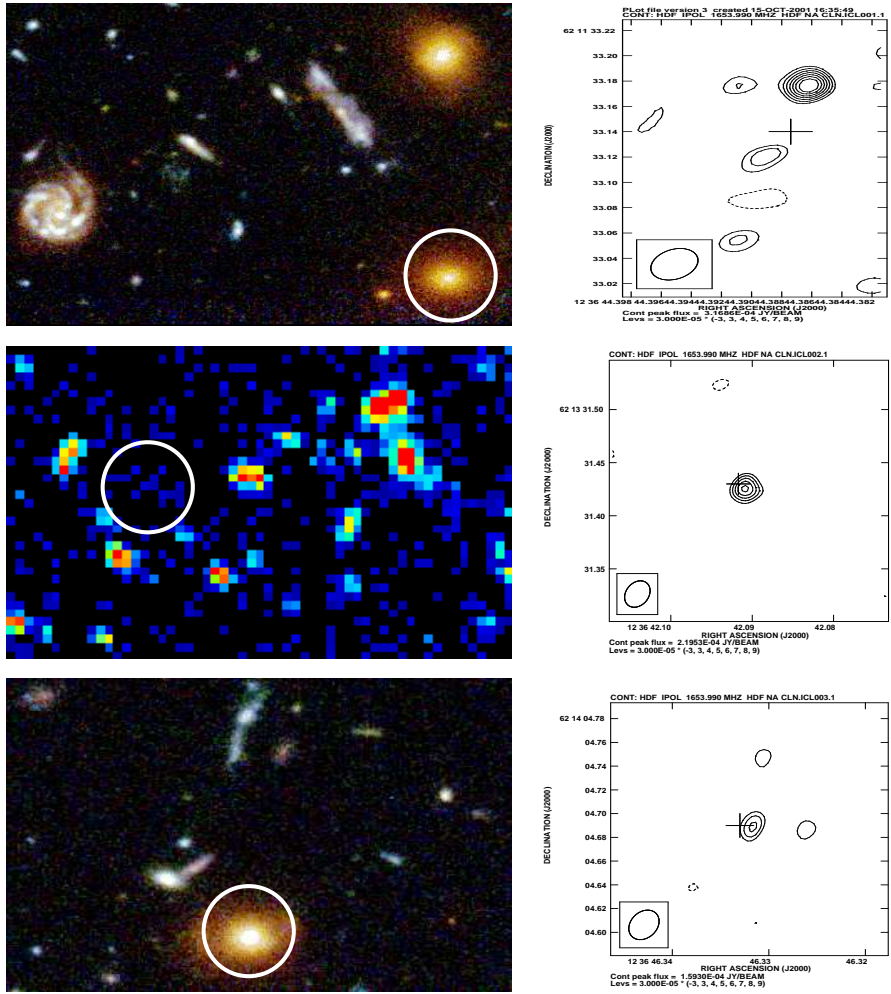}
\caption{EVN detections in the HDF: the distant
  z=1.01 FRI (top), the z=4.4 dusty obscured starburst hosting a
  hidden AGN (middle) and the
  z=0.96 AGN (bottom). Crosses represent the MERLIN-VLA source positions.}
\label{evn_im}
\end{figure}

\section{The role of VLBI} 

In principle extremely sensitive global VLBI observations have sufficient
resolution to distinguish between AGN and starburst activity in
these optically faint radio sources. A taste of what may be
possible is provided by the recent EVN observations of the Hubble Deep
Field. 

\subsection{Deep Field VLBI observations of the HDF} 
\label{EVN}

On 12-14 November 1999 the EVN conducted the first pilot VLBI
``blank field'' observations of the radio sky. The field chosen was the
HDF-N -- an area that is just about as empty and undistinguished as
the radio sky gets.  The brightest source in the $\sim 2$ arcminute
radial field of view was an FR-I radio galaxy with a total WSRT 1.4~GHz
flux density of $\sim 1.6$~mJy. 

The data were recorded at a rate of 256 Mbits/sec for 32 hours -- a
sustained capability that is unique to the EVN (and has recently been
extended to 512 Mbits/sec).  Observing in external phase-reference
mode, a total of $\sim 14$ hours of ``on-source'' data were collected.
With a resolving beam area 1 million times smaller than the WSRT HDF-N
observations, the EVN surveyed an area of about 12 square arcminutes.
Six HDF-N radio sources were thus targeted simultaneously (using
wide-field imaging techniques -- see Garrett et al. 1999). The final
naturally weighted images have an r.m.s. noise level of $\sim
33~\mu$Jy/beam -- much larger than that expected from thermal noise
considerations ($\sim 11~\mu$Jy/beam). The images are probably limited
by the inclusion of poorly calibrated or completely corrupt data -
difficult to identify in this case, and not uncommon with external
phase-calibration during observations made at solar maximum. 

Nevertheless, the EVN simultaneously detected three radio sources above
the $165~\mu$Jy ($\sim 5\sigma$) limit, in the inner part of HDF-N
region (see Fig~\ref{evn_im}). These include: VLA~J123644+621133 (a
$z=1.013$, low-luminosity and extremely distant FR-I radio source which
is resolved by the EVN into a core plus hot-spots, the latter being
associated with the larger scale radio jet), VLA~J123642+621331 (a dust
enshrouded, optically faint -- $I < 25^{m}$, $z=4.424$ starburst system
-- Waddington et al.  1999) and the faintest detection,
VLA~J123646+621404 (a face-on spiral galaxy at $z=0.96$ with a total
EVN flux density of $180~\mu$Jy/beam).

The diversity of optical type is interesting but the real surprise is
the detection of a radio-loud AGN in the dust obscured, $z=4.4$,
starburst system. This argues that at least some fraction of the
optically faint radio source population harbour hidden AGN (this may be
similar to the same obscured population detected by Chandra). These AGN
powered systems might be quite difficult to detect with SCUBA, if
the dust temperatures are higher than that associated with pure star
forming systems. In any case, the detection of this system highlights
the use of VLBI as a powerful diagnostic -- able to distinguish in
principle (via brightness temperature arguments) between radio emission
generated by nuclear starbursts and AGN.

\begin{figure}[h]
\vspace{11.5cm}  
\includegraphics{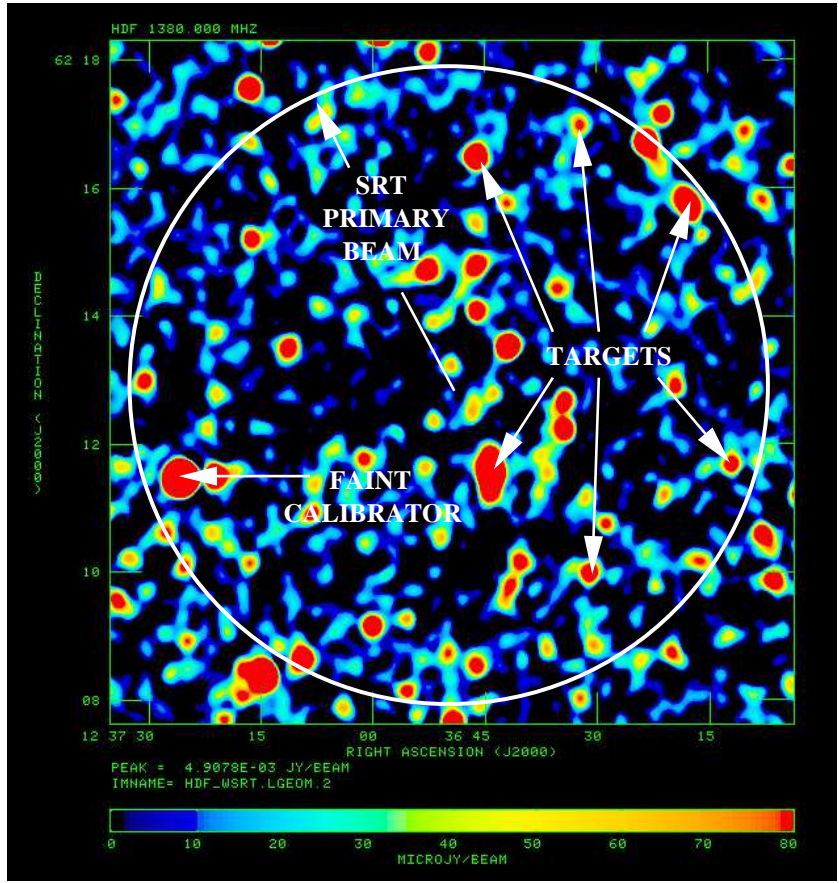}

\caption{The SRT 64-m primary beam at L-band, superimposed upon 
  the crowded WSRT 1.4~GHz image of the HDF region. Around 100 radio sources
  (brighter than 40 microJy/beam) and potential VLBI targets are
  enclosed within the SRT primary beam. The faint, ``in-beam''
  calibrator that permits accurate and continuous phase calibration to
  be applied to the other target sources is also shown.  
}
\label{inbeam} 
\end{figure}

\section{Future Prospects for Deep Field VLBI Studies}

The EVN observations of the HDF suggest that deep, wide-field VLBI
studies are not only possible, but in principle they can deliver
important astronomical results. So far we have only scratched the
surface. In a sense, we are just beginning to appreciate the fact, that
VLBI has reached a sensitivity level where we can expect to detect many
discrete radio sources in a single field of view ({\it irrespective} of
where you point the telescopes!). This is quite a departure from the
traditional VLBI ``postage stamp'' approach of observing specific, singular
targets, often chosen from flux limited catalogues of the very
brightest and most compact radio sources in the sky. 

\subsection{Short-term developments} 

In the short-term continuum VLBI sensitivity is set to continue its
steady improvement -- three new telescopes are expected to come on-line
(including the 64-m Sardinia Radio Telescope, see Grueff these
proceedings) and various upgrades are already well advanced (e.g. the
replacement of the surface of the Lovell 76-m). Recent EVN test
observations have demonstrated that the new 2-head, 512 Mbit/sec data
recording system can reach thermally limited r.m.s. noise levels of
$\sim 20$~microJy/beam with (on-source) integration times of only $\sim
2.5$ hours.  There is every reason to believe that much longer
observations employing ``in-beam'' phase-calibration techniques (rather
than external phase-calibration) can also attain thermally limited
noise levels. For example, currently a 24 hour observation of a
carefully selected deep field should attain a r.m.s. noise level of
$\sim 7$~microJy/beam. Similar noise levels can be achieved with a
global VLBI array. At these sort of flux density levels, the radio sky
becomes so crowded that it makes sense to image out a large fraction of
the primary beam of the largest individual antennas in the array. The
anticipated introduction of extremely fine spectral resolution and
sub-second integration times at the EVN Data Processor at JIVE, will
make it possible to do just this.  Dozens of faint sub-mJy and microJy
radio sources will thus be simultaneously targeted and imaged with
milliarcsecond resolution, full uv-coverage and microJy sensitivity.

The ``postage stamp'' era of VLBI will thus be consigned to the
astronomical dust bin, and with it, the current limitations of VLBI to
conduct unbiased, general sky surveys.
Fig.~\ref{inbeam} highlights the type of VLBI, deep field observing
philosophy that might be employed. The ``in-beam'' calibrator is chosen
to be bright enough to provide adequate SNR in the phase solutions but
not so bright as to impose dynamic range limitations on the surrounding
field (see Garrett 2000b for a more detailed explanation). 

\subsection{Longer-term expectations} 

\begin{figure}[h]
\vspace{6.0cm}  
\includegraphics{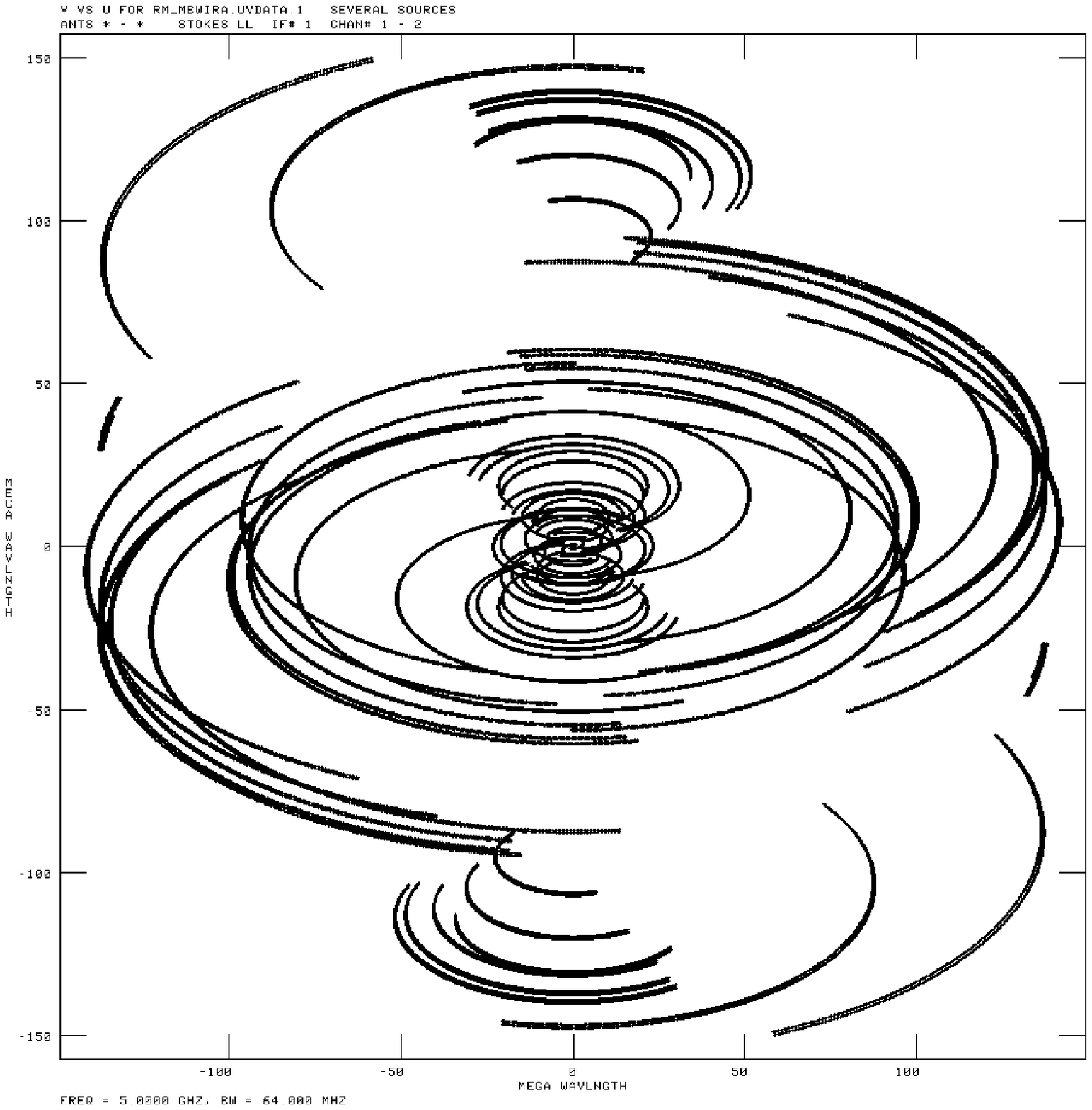}
\includegraphics{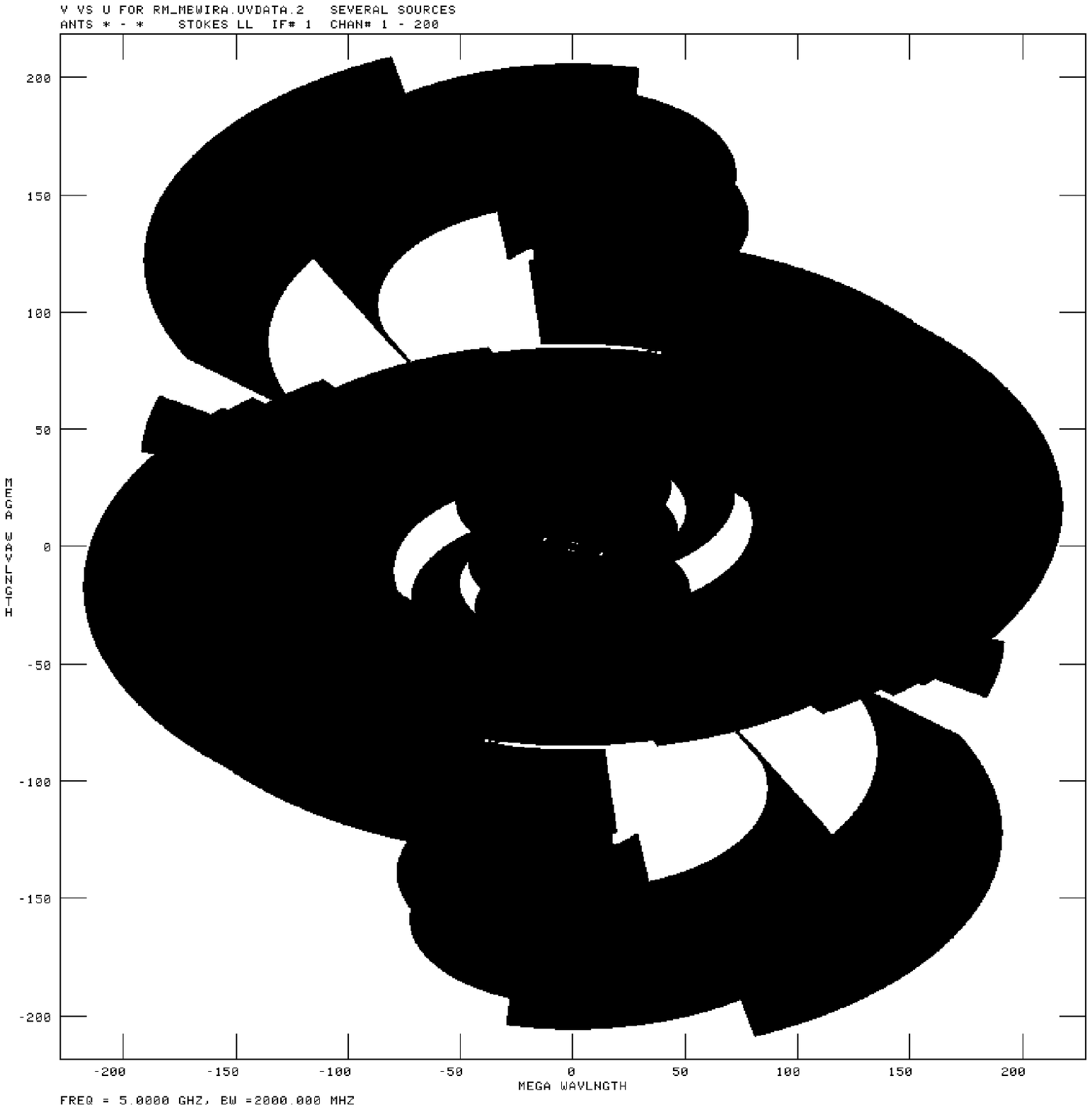}
\caption{Left: The current uv-coverage of the EVN at $\lambda6$~cm for a source
  located at $\delta=30^{\circ}$. Right: the extended (almost full)
  uv-coverage of the $e$EVN for the same source, assuming a total
  bandwidth of 2~GHz per polarisation.}
\label{uvcov} 
\end{figure}

PC disk-based ``recording'' systems (e.g. MkV, Whitney 2001 and PC-EVN,
Parsley 2001) are expected to replace the aging MkIV and VLBA tape
recorders over the next few years. Initially they provide the same
level of economy, capacity and recording data rates as current tape
based systems but with a much reduced level of maintenance and a much
higher level of reliability. In addition, the capacity of these
disk-based systems is also expected to increase rapidly, as they take
advantage of the continuing commercial development of PCs and their
peripheral storage devices.  Commercially operated optical fibre
networks also appear capable of distributing VLBI data from telescopes
to a central correlator in real-time. Tests are already being
conducted, both within Europe, the US and between the two continents
(see Schilizzi these proceedings \& Schilizzi 2002).  In any
case, the expectation is that VLBI will be able to utilise several GHz
of bandwidth on the same time-scale as the EVLA and e-MERLIN begin
operation i.e. before the end of the current decade.  Employing several
GHz of bandwidth has of course implications down the line: a more
capable correlator will be required, perhaps not very different to the system 
now being designed for the EVLA and e-MERLIN. VLBI data acquisition electronics
will also need to be completely redesigned, and broad-band receivers
must be introduced at telescopes across the network.  A new instrument
will thus emerge (currently dubbed $e$EVN) that will deliver
substantial benefits: sub-microJy noise levels and almost full
uv-coverage at $\lambda 6$~cm (see Fig.~\ref{uvcov}).

\subsection{New Deep Field VLBI Science} 

\begin{figure}[h]
\vspace{8cm}  
\includegraphics{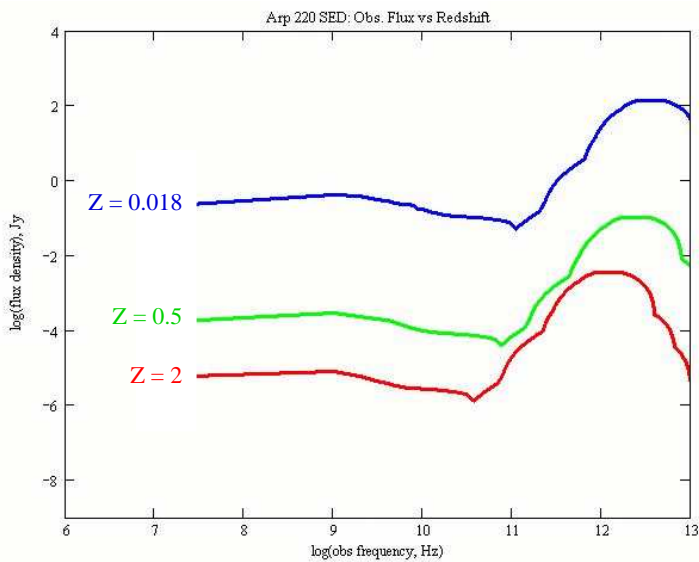}

\caption{The SED of Arp 220 as observed and then projected ($\Omega_{{\rm m}}=0.3$, $\Omega_{\Lambda}=0.7$, $H_{{\rm
      0}}=70$~km/sec/Mpc) to z=0.5 and z=2.}
\label{arp220} 
\end{figure}

At these sub-microJy noise levels, it should be possible to begin to
detect not only traditional VLBI targets such as low-luminosity AGN, SN
(SNR) etc., but also cosmologically distant, nuclear starburst
galaxies, such as those now observed (but typically not well resolved)
by connected arrays in the Hubble Deep Field. Fig.~\ref{arp220} shows
how the radio and sub-mm flux density of Arp 220 changes with redshift.
In nuclear starbursts (such as Arp 220) a large fraction of the radio
emission is contained within a region of intense star formation,
usually less than a kpc across. At cosmological distances 1 kpc
corresponds to an angular diameter of $\sim 0.1$ arcsecond on the sky,
a scale size that is ideally suited to a 
combined $e$EVN and e-MERLIN array.

With a star formation rate (SFR) of $\sim 100$~M$_{\odot}$/yr,
Fig.~\ref{arp220} shows that Arp 220 could in principle be detected by
the $e$EVN out to cosmologically interesting redshifts. Since many of
the galaxies in the HDF have inferred SFR that are more than an order
of magnitude larger than Arp 220, the radio structure of galaxies at
even earlier epochs may be probed. It is not clear what to expect but
the view is expected to be dramatic, as massive star formation (and its
violent ``radio loud'' aftermath -- SN, SNR and GRBs) rampages
unchecked through still-forming galaxies. A detailed study of the radio
structure of such distant systems will probably require the nanoJy
sensitivity of a high resolution, next generation radio instrument, such
as the Square Km Array (SKA). Deep radio continuum images can tell us
how star formation is distributed through a given system, providing
clues to the nature of the galaxy formation process itself and the
importance of galaxy interactions and mergers, not to mention the
inter-relationship between AGN activity and star formation.  But in the
meantime, the $e$EVN with baselines on the scale of several thousand
km, GHz of bandwidth, ``fantastic'' input and output data rates, and of
course microJy sensitivity, can make a useful start in this area. The
$e$EVN will also form a natural test-bed for the SKA, encountering many
of the same problems and initial limitations. The next few years
promise to be exciting and challenging, for radio astronomy in general
but VLBI in particular.

\acknowledgments

Several of the results shown here were made in collaboration with
several colleagues. I'd particularly like to acknowledge contributions
made by Ger de Bruyn, Simon Garrington and Tom Muxlow. I'd also like to
thank the organisers of the meeting for their support, {\it pazienza}
(particularly with regard to the production of this paper) and
the memorable entertainment at the conference dinner!

%

%
%
\end{document}